# An Enhanced Electrocardiogram Biometric Authentication System Using Machine Learning

Ebrahim Al Alkeem[1], Song-Kyoo Kim[2], Chan Yeob Yeun[1,2], M. Jamal Zemerly[1], Kin Poon[3], Gabriele Gianini[3,4] and Paul D. Yoo[5,6]

[1]Department of Electrical Engineering and Computer Science (EECS), Khalifa University of Science and Technology, Abu Dhabi, UAE
[2]Center for Cyber-Physical Systems (C2PS), Khalifa University of Science and Technology, Abu Dhabi, UAE
[3]Emirates ICT Innovation Centre (EBTIC), Khalifa University of Science and Technology, Abu Dhabi, UAE
[4]Dipartimento di Informatica "Gianni Degli Antoni", Università degli Studi di Milano, Italy
[5]CSIS, Birkbeck College, University of London, Malet Street, London, WC1E 7HX, United Kingdom
[6]Cranfield School of Defence and Security, Defence Academy of the United Kingdom, Shrivenham, SN6 8LA, United Kingdom

Corresponding author: Chan Yeob Yeun (e-mail: chan.yeun@ku.ac.ae)

**ABSTRACT** Traditional authentication systems use alphanumeric or graphical passwords, or token-based techniques that require "something you know and something you have". The disadvantages of these systems include the risks of forgetfulness, loss, and theft. To address these shortcomings, biometric authentication is rapidly replacing traditional authentication methods and is becoming a part of everyday life. The electrocardiogram (ECG) is one of the most recent traits considered for biometric purposes. In this work we describe an ECG-based authentication system suitable for security checks and hospital environments. The proposed system will help investigators studying ECG-based biometric authentication techniques to define dataset boundaries and to acquire high-quality training data. We evaluated the performance of the proposed system and found that it could achieve up to the 92% identification accuracy. In addition, by applying the Amang ECG (*amgecg*) toolbox within MATLAB, we investigated the two parameters that directly affect the accuracy of authentication: the ECG slicing time (sliding window) and the sampling time period, and found their optimal values.

**INDEX TERMS** authentication, biomedical signal processing, electrocardiogram signal (ECG), machine learning, multi-variable regression

## I. INTRODUCTION
Biometric authentication is replacing typical identification and access control systems to become a part of everyday life [1, 2]. The electrocardiogram (ECG) is one of the most recent traits to be explored for biometric purposes [3, 4]. ECGs report electrical conduction through the heart and can be used to recognize specific individuals [5]. The utilization of ECGs as a biometric trait was first proposed in a 1977 US military report [6]. Although much progress has been achieved over the last decades, many challenges remain to be overcome [5], including data acquisition, pre-processing for data enhancement, the assignment of authentication categories. However, the recent development of deep learning (DL) and other machine learning (ML) classification techniques [7] open new perspectives to this approach to authentication. ML techniques have recently been used to construct a verification model for identification based on live ECG data [8-11]. ML is a subfield of Artificial Intelligence: ML algorithms build a mathematical model based on training data; typical models are regression (predictions) models and decisions models (e.g. classification and pattern recognition) [12]. The diverse applications of ML include the analysis of videos, images, and sounds [13], as well as ECG data [11, 14-15].

ECG research covers a wide range of disciplines with different requirements. Medical engineers set up electrocardiographs for the collection of rich ECG data [5, 16-17], whereas electrical engineers use simpler sensors to detect ECG signals [18-21]. In a previous study [22] we defined three use cases that influence the setup of an ECG-based authentication system focusing the attention on aspects of the system that are relevant for external users [23]. The three use cases are security checks (SCK), hospitals (HOS) and wearable devices (WD). The analysis of these three use cases helps researchers in the field of biometric authentication to understand the conditions and setup the requirements for each scenario [22]. In this article, we consider the SCK and HOS use cases in more detail. In a typical SCK scenario, biometric authentication based on a simple ECG scan, would take place at a security checkpoint of an entrance to a building and would identify employees and visitors, while excluding unknown persons [22]. In contrast, the HOS use case involves complex medical equipment which collects detailed ECG data during the training and testing phases. This requires a longer sampling time period and multiple leads to attach a person to gather the data [22].

Our proposed authentication system uses multi-variable regression to break down the dataset into smaller subsets, then builds a decision tree (DT) model based on these variables to predict target values [24-25].

With respect to competing techniques, such as DL techniques, we consider the following. The convoluted nature of popular deep-structured machine learning implies lack of transparency and interpretability. The knowledge obtained by more interpretable learners (such as decision trees) is critical in biometric software design. The deep learning methods [26-29] could be applied but the DT is more flexible for the input size of the samples and the changes in sampling frequency.

We use the time-sliced ECG method to build the training and testing datasets [22] and also investigate the optimum sliding window size. Time-sliced ECG data provides a sufficient number of samples and each sliced dataset can be used as the input for ML training. Although time-sliced ECG data offers sufficient flexibility to mix with other training inputs, we have used this source of data on its own. Previous research showed that the performance of ML is dependent on the ECG slicing time [7-10], we have therefore investigated this relationship. The minimum heartbeat interval within a typical heart rate [29] was chosen as the sliding window width. Representative time-sliced ECG data are shown in Fig. 1. We used the Amang ECG (*amgecg*) toolbox in MATLAB for ECG time slicing and to build the training input for the regression approach [22].

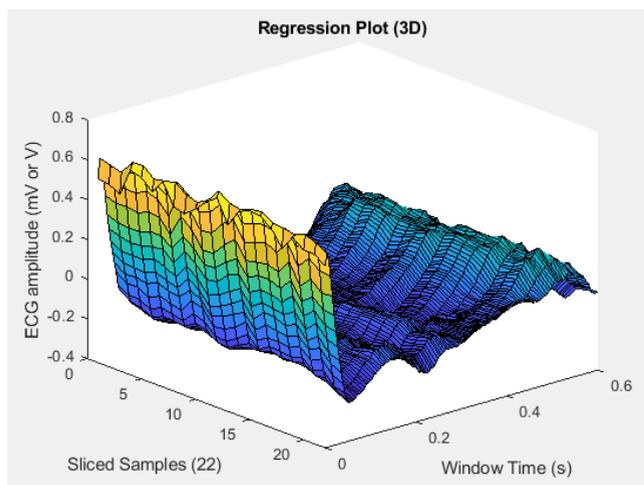

**FIGURE 1. ECG time slicing with R-peak anchoring [31].**

This paper is organized into four sections. Section II describes the new ML-based authentication system and evaluates its classification performance [32]. Section III considers the two most relevant parameters (slicing time and sampling time period) that directly affect authentication performance: their relationship and impact are evaluated. Finally, the new authentication system and its contributions are summarized in Section IV.

## II. ECG-BASED AUTHENTICATION USING ML

This section describes the ECG-based authentication system using regression as a ML technique that particularly complements the security check, or SCK, use case. The sampling time period for the testing (validation) phase is relatively short: less than 20s. The system can identify the unknown entity within this short interval [22].

### A. SECURITY CHECK CASE: EXPERIMENTAL SETUPS

Several ML approaches could be used to develop a regression model, however our previous studies showed that the DT method achieves the best performance with time-sliced ECG data [22]. We confirmed this by comparing the performance of the decision tree (a fine tree, with a rich structure, i.e. many nodes) and support vector machine (SVM) methods and the results are shown in Table I. Based on these results, we selected the DT-based regression method for our authentication system. The performance comparison between DT and SVM have been analyzed by the *Regression Leaner* function in Matlab. The values in Table I is automatically generated by the Matlab function during training the dataset.

TABLE I
PERFORMANCE COMPARISON BETWEEN DECISION TREES (DT) AND SUPPORT VECTOR MACHINES (SVM)

|  | DT (Fine Tree) | SVM (Fine Gaussian) |
|---|---|---|
| RMSE[1] | 0.05817 mV | 0.06011 mV |
| MAE[2] | 0.00339 mV | 0.00361 mV |
| Training Time | 5.96 s | 7554.3 s |

We therefore applied this method to the SCK use case based on 90 ECG data samples arbitrary collected in a HOS environment [22]. An ECG-based authentication system can be used to identify employees and exclude unknown persons assuming that employees have registered their identities and the ECG data are stable enough during both the training and testing phases. Out of the 90 samples used to construct the dataset for this experiment, 63 were sourced from the PhysioBank database [31, 33] and 27 from the Diabetes Complications Research Initiative [34]. The data from the PhysioBank [31, 33] have been downloaded and transforming into Matlab formats by using the WFDB toolbox [35] and the data from the other source [34] are originally supported as the Matlab format. Ten additional samples (indicated hereafter as 'unknown') were randomly selected from the dataset using the same sources but in different sampling time periods [31, 33-34] and added during the testing phase [22]. Pre-processing for the HOS use case is still necessary even when dealing with the SCK scenario because the sources are not originally from the security check.

The authentication process began with use-case categorization and pre-processing, before training the dataset. All pre-processing steps recommended in our previous study [22] were applied (including baseline drift adjustment, power line interference (PLI) noise adjustment and checking the flipping signal) because the data were sourced from medical equipment (HOS use case). Although the pre-process for ECG signals by using various techniques has been widely studied [36-39], our system only follows the basic standard methods which have been suggested from the ML framework [22]. The polynomial curve fitting for the baseline drift adjustment and the Fourier Transform for the PLI noise cancellation have been applied in this research [22]. The ECG data were collected at two different times, resulting in two datasets, which were defined as the training and testing sets. The ECG data were trained using the DT regression method (Fig. 2).

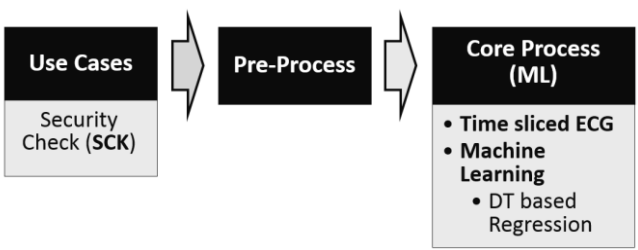

**FIGURE 2.** The training process for the SCK use case.

Although many different measures and techniques are cited in the biometric literature for feature selection filters, we used mutual information in DT model as a measure to score and rank the features. Mutual information was proposed by Shannon

---

[1] RMSE: Root Mean Square Error
[2] MAE: Mean Absolute Error

[40] as part of this information theory, and was based in the concept of entropy, used to quantify the irreducible complexity of random signals (below which no lossless compression is possible). The entropy $H$ of a random variable $x$ with a probability mass function $p(x)$ is

$$H(x) = -\sum_{x \in X} p(x) \log_2 p(x) \quad (1)$$

where $X$ is a set of all possible outcomes of $x$. From the above definition derive the definitions of: (i) conditional entropy,

$$H(x|y) = -\sum_{x \in X}\sum_{y \in Y} p(x,y) \log_2[p(x|y)] \quad (2)$$

quantifying the entropy of a random variable $x$ conditional upon the knowledge of another random variable $y$; and (ii) mutual information,

$$I(x;y) = \sum_{x,y} p(x,y) \log_2 \frac{p(x,y)}{p(x)p(y)} = H(x) - H(x|y), \quad (3)$$

denoting the amount of information gained about $y$ as a result of knowing $x$. In terms of our mutual information theoretic feature selection filter, the random variables $x$ and $y$ can be used to represent features and class labels: $x$ can be used to denote a feature within the dataset, and $y$ can be used to denote a class label. The information theoretic measures help quantifying how likely the machine learner is to correctly predict the class label, for any given instance, as a result of learning a given feature. Once mutual information is estimated, one can rank the features based on their capability to predict the target and selected accordingly. We point out that this approach to feature selection inherently relies on the assumption that classifier performance is linked to the amount of mutual information shared between the class label and a feature, *i.e.* the greater the number of high-ranking features selected for classifier training, the better the classifiers perform in terms of correctly identifying the class label for any given instance. However, this approach disregards the potential for better subsets to exist, comprising features that are not sequentially ranked in terms of their mutual information values.

Computing the equations from (1) to (3) requires the knowledge of probability measures $p(x)$, $p(y)$ and $p(x,y)$. Since these quantities are frequently unknown *a priori* for any given datasets, we invoke a widely-used histogram-based approach [41] to estimate the probability distribution. Generally *histogramming* is known to introduce estimation bias due to sensitivity to bin size.

The sliced ECG time (sliding window) for this experiment is 0.6 s which is equivalent to the interval between heartbeats at a typical rate of 100 beats per minute [30]. However, the slice time can be changed, and may therefore affect the authentication performance, and this relationship is discussed in Section III. The detailed process flow for the training and testing phases is summarized in Fig. 3.

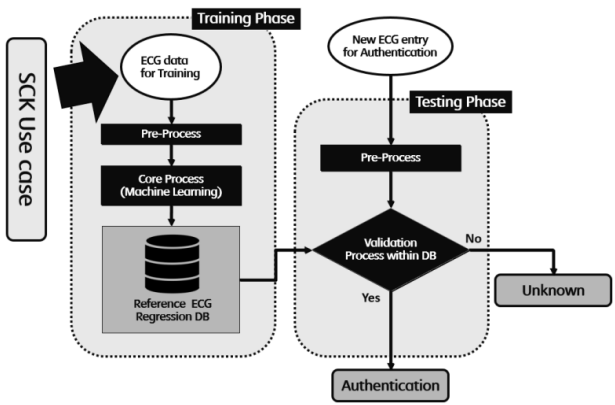

FIGURE 3. Process flow for ECG-based authentication in the SCK use case.

The training phase generates the reference regression functions for each sample (i.e., entity) using the DT technique and stores all functions as a database. This database is then used to compare the ECG data when new ECG data are detected during the testing phase. The sampling time period for the training data was set to 50s. The sampling time period for the testing data should be shorter due to the properties of the SCK use case category. The detection of the ECG testing data should be faster because the SCK use case considers a scenario in which employees are entering a company building. Accordingly, the sampling time period for testing was set to 15s. The core process generates reference regression functions for each set of ECG training data. Some of the trained reference functions are shown in Fig. 4 and these can be compared with the ECG testing data without fixing the sampling frequency.

### B. EXPERIMENTAL RESULTS

Authentication performance was evaluated by means of standard metrics [32] such as accuracy and recall a.k.a. sensitivity). Given the characteristics of the SCK use case, the authentication process focuses on the detection of unknown entities, thus the recall is defined as the ratio of the detected unknown and the total of unknown entities. We also applied a data quality measure based on the mean square error (MSE) before starting to detect the testing ECG data. The experiment was performed 150 times using 100 samples, with a sampling time period of 15s for the authentication testing and the confusion matrix. Notably, 28 out of the 150 ECG datasets (17.61%) were rejected because they did not meet the data quality criteria [22]. The resulting dataset consisted in 122 samples. The confusion matrix resulting from the experiment is shown in Table II.

TABLE II
AUTHENTICATION CONFUSION MATRIX FOR SCK USE CASE I

| 122 | | Actual ECG data | |
|---|---|---|---|
| | | Known | Unknown |
| Predicted ECG data | Known | 84 | 2 |
| | Unknown | 30 | 6 |

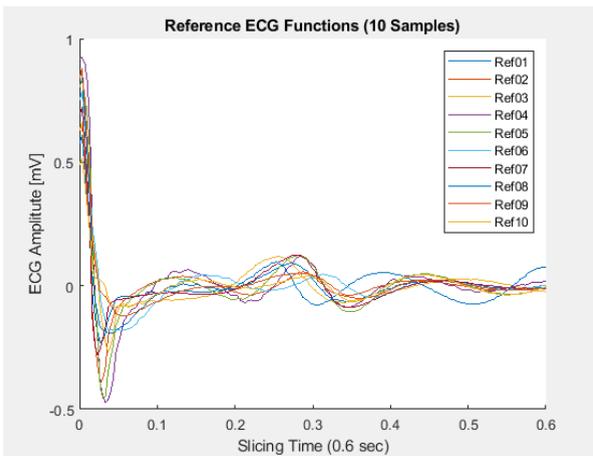

**FIGURE 4.** The reference ECG regression functions for the SCK use case.

The values of the performance metrics are the following: overall accuracy 90/122 (73.77%); recall (proportion of unknown actually detected) 6/8 (75%).

The acceptance criteria for validation could be strengthened based on the quality of the training data (according to the upper control limit of the MSE). When these higher data quality criteria were used, the samples were reduced to 82. The corresponding confusion matrix is shown in Table III.

TABLE III
AUTHENTICATION CONFUSION MATRIX FOR SCK USE CASE II

| 82 | | Actual ECG data | |
|---|---|---|---|
| | | Known | Unknown |
| Predicted ECG data | Known | 76 | 5 |
| | Unknown | 1 | 0 |

The accuracy of this biometric authentication system was 76/82 (92.7%). However, the recall was 0. Notably, the values could vary because the ECG testing data were randomly selected for each trial to make the SCK use case more realistic.

## III. SLICING AND SAMPLING TIME PERIOD DEPENDENCIES

Some ML performance measures depend on the ECG slicing time (sliding window). We gathered 70 samples from the various sources discussed above [31, 33-34] and the HOS use case was addressed to find a relationship between the authentication performance and two key parameters: the sliding window size (i.e., ECG slice time) and the ECG data sampling time period. The relationship between the slicing time and authentication accuracy is shown in Fig. 5, revealing that the optimal slicing time is approximately half the average interval between heartbeats (0.4s).

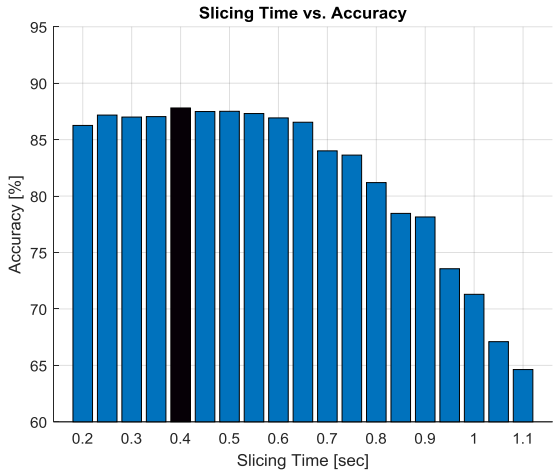

**FIGURE 5. Authentication accuracy based on slicing time.**

We also investigated the relationship between sampling time period and authentication accuracy (Fig. 6). Although there was no clear relationship between these parameters, the optimal sliding time was 37 seconds in our experiment. The optimal slicing and sampling time periods may not remain the same if different datasets are used. However, our experiments clearly demonstrated that optimal values for these parameters exist and can be used to improve performance.

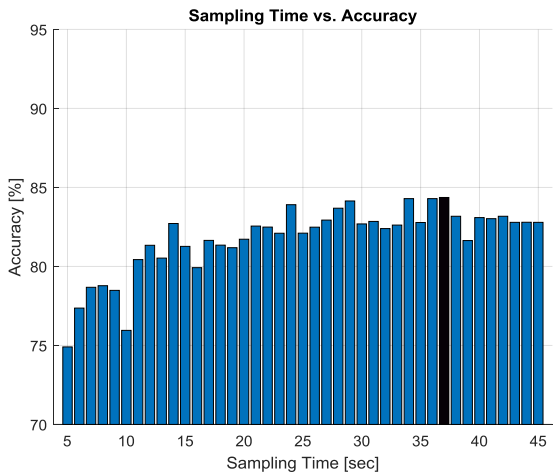

**FIGURE 6. Authentication accuracy based on sampling time period.**

## IV. CONCLUSION

In this paper, we proposed an enhanced ECG-based biometric authentication system for the SCK use case in which a regression-based interpretable ML approach was used to define the dataset boundaries and to acquire good-quality training data. We trained on a total of 90 ECG data samples to generate the reference function database. The reference function for each ECG data entity (i.e., identification) was then generated using a mutual-information-based DT regression approach. The authentication performance of the proposed system was evaluated not only with a confusion matrix but also by using the *amgecg* toolbox in MATLAB to analyze two key parameters: the ECG slicing time (sliding window) and the sampling time period. We found that a sliding window of 0.4s achieved the best performance and that the optimal sampling duration is 37s. In conclusion, using these optimized parameters, the proposed authentication system is able to achieve accurate results.

# APPENDIX

The *amgecg* toolbox v.0.5 [22] and WFDB toolbox v0.10.0 [35] were used to design our authentication system. The corresponding MATLAB codes are available on GitHub (https://github.com/amangkim/ECGRegreSec) for users to try the demonstrations.

# REFERENCES

[1] F. Agrafioti, J. Gao, Hatzinakos, D. Heart Biometrics: Theory, Methods, and Applications. In Biometrics; Yang, J., Ed.; InTech: Rijeka, Croatia, pp. 199–216, 2011.
[2] A. K. Jain and et al., Nandakumar, K. Introduction to Biometrics; Springer Science+Business Media, LLC, New York, NY, USA, 2011.
[3] F. Agrafioti, F. M. Bui and et al., Secure Telemedicine: Biometrics for Remote and Continuous Patient Verification. J. Comput. Netw. Commun. 924791, 2012.
[4] M. Li, and S. Narayanan, Robust ECG Biometrics by Fusing Temporal and Cepstral Information. In Proceedings of the 2010 20th International Conference on Pattern Recognition (ICPR), Istanbul, Turkey, 23-26 August, pp. 1326-1329, 2010.
[5] J. R. Pinto, J. S. Cardoso, and et al., "Towards a Continuous Biometric System Based on ECG Signals Acquired on the Steering Wheel", Sensors 2017, vol. 17 no. 10, 14 pages, 2017.
[6] G. E. Forsen, M. R. Nelson and et al., Personal Attributes Authentication Techniques; Technical Report; Pattern Analysis and Recognition Corporation, Rome Air Development Center: Rome, NY, USA, 1977.
[7] Y. Xin, L. Kong, Z. Liu, Y.Chen, Y. Li, H. Zhu, M. Cao, H. Hou, and C. Wang, "Machine Learning and Deep Learning Methods for Cybersecurity", IEEE Access, vol. 6, pp. 35365-35381, 2018.
[8] Q. Zhang, D. Zhou and X. Zeng, "HeartID: A Multiresolution Convolution Neural Network for ECG-Based Biometrics Human Identification in Smart Health Applications", IEEE Access, vol. 5, pp. 11805-11816, 2017.
[9] J.R. Pinto, J.S. Cardoso and A. Lourenco, "Evolution, Current Challenges, and Future Possibilities in ECG Biometrics", IEEE Access, vol. 6, pp. 34746-34776, 2018.
[10] E.J.S. Luz, G.J.P. Moreira, L.S. Oliveira, W.R. Schwartz, and D. Menotti, "Learning Deep Off-the-Person Heart Biometrics Representations", IEEE Transaction on Information Forensics and Security, vol. 13, no. 5, pp. 1258-1270, 2018.
[11] H. Kim and S.Y. Chun, "Cancelable ECG Biometrics Using Compressive Sensing-Generalized Likelihood Ratio Test", IEEE Access, vol. 7, pp. 9232-9242, 2019.
[12] F. H. Bennett and et al., "Automated Design of Both the Topology and Sizing of Analog Electrical Circuits Using Genetic Programming" Artificial Intelligence in Design 96 Springer, Dordrecht. pp. 151-170, 1996.
[13] F. Camastra, A. Vinciarelli, Machine Learning for Audio, Image and Video Analysis 2nd Ed., Springer, New York, NY, 2015.
[14] A. Minchole and B. Rodriguez, "Artificial intelligence for the electrocardiogram", Nature Medicinevolume, vol. 25, pp. 22-23, 2019.
[15] E. Tatara and A. Cinar, "Interpreting ECG data by integrating statistical and artificial intelligence tools", IEEE Engineering in Medicine and Biology Magazine, vol. 21, no. 1, pp. 36-41, Jan.-Feb. 2002.
[16] H. J. Kim and J. S. Lim, "Study on a Biometric Authentication Model based on ECG using a Fuzzy Neural Network", 2018 IOP Conf. Ser.: Mater. Sci. Eng. 317, 10 pages, 2018.
[17] M. Sansone, R. Fusco and et al., "Electrocardiogram Pattern Recognition and Analysis Based on Artificial Neural Networks and Support Vector Machines: A Review", Journal of Healthcare Engineering, vol. 4, no. 4, pp. 465-504, 2013.
[18] E. Saddik and et al., "Electrocardiogram (ECG) Biometric Authentication", U. S. Patent 9,699,182 B2, Jul. 4, 2017.
[19] S. Y. Chun, J.-H. Kang and et al., "ECG Based User Authentication For Wearable Devices Using Short Time Fourier Transform", in Proc. 39th IC-TSP, Vienna, Austria, 2016, pp. 656-659.
[20] A. F. Hussein, A. K. AlZubaidi and et al., "An IoT Real-Time Biometric Authentication System Based on ECG Fiducial Extracted Features Using Discrete Cosine Transform", [Online]. Available: https://arxiv.org/abs/1708.08189
[21] B.E., Manjunathswamy, A. M., Abhishek and et al., "Multimodal Biometric Authentication using ECG and Fingerprint", International Journal of Computer Applications, 111:13, pp. 33-39, 2015.
[22] S. -K. Kim, C. Y. Yeun and et al., "A Machine Learning Framework for Biometric Authentication using Electrocardiogram", IEEE Access **7**, pp. 94858-94868, 2019.
[23] usability.gov, Use Cases, [Online]. Available: https://www.usability.gov/how-to-and-tools/methods/use-cases.html, Accessed on Feb. 1, 2019
[24] P. Gupta, "Decision Trees in Machine Learning", Towards Data Science, [Onlne] Available: https://towardsdatascience.com/decision-trees-in-machine-learning-641b9c4e8052, Accessed on Feb. 1, 2019.
[25] J. R. Quinlan, Induction of Decision Trees. Mach. Learn. 1, 1, pp. 81-106, 1986.
[26] M. Gerven1, S. Bohte, "Artificial Neural Networks as Models of Neural Information Processing", Front. Comput. Neurosci., 19 December 2017 [Online] Available: https://www.frontiersin.org/articles/10.3389/fncom.2017.00114/full
[27] L. Yann, "LeNet-5, convolutional neural networks", [Online] Available: http://yann.lecun.com/exdb/lenet/, Accessed on Feb. 1, 2019
[28] W. Yu, K. Yang and et al., "Visualizing and Comparing Convolutional Neural Networks", [Online]. Available: https://arxiv.org/abs/1412.6631, Accessed on Jul. 1, 2019.
[29] R. D. Labati, E. Munoz and et al., Deep-ECG: Convolutional Neural Networks for ECG biometric recognition, Pattern Recognition Letters [Online] https://doi.org/10.1016/j.patrec.2018.03.028, Accessed on Jul. 1, 2019.
[30] American Heart Association, "All About Heart Rate (Pulse)", [Online], Available: https://www.heart.org/en/health-topics/high-blood-pressure/the-facts-about-high-blood-pressure/all-about-heart-rate-pulse, Accessed on Feb. 1, 2019.
[31] A. Taddei, G. Distante and et al., The European ST-T Database: standard for evaluating systems for the analysis of ST-T changes in ambulatory electrocardiography. European Heart Journal 13: 1164-1172, 1992.
[32] T. Fawcett, "An Introduction to ROC Analysis", Pattern Recognition Letters. 27:8, pp. 861-874, 2005.
[33] A. L. Goldberger, L. A. N. Amaral and et al., PhysioBank, PhysioToolkit, and PhysioNet: Components of a New Research Resource for Complex Physiologic Signals, Circulation vol. 101, no. 23, pp. e215-e220, June, 2000. [Online]. Available: http://circ.ahajournals.org/content/101/23/e215.full.
[34] M. H. Imam, C. K. Karmakar and et al, "Detecting Subclinical Diabetic Cardiac Autonomic Neuropathy by Analyzing Ventricular Repolarization Dynamics", IEEE J. Biomedical and Health Informatics, vol. 20 no. 1, pp. 64-72, 2016.

[35] Silva, I, Moody, G. "An Open-source Toolbox for Analysing and Processing PhysioNet Databases in MATLAB and Octave." Journal of Open Research Software 2(1):e27 [http://dx.doi.org/10.5334/jors.bi] ; 2014 (September 24).
[36] Y. -W Bai, W. -Y. Chu and et al., "Adjustable 60Hz noise reduction by a notch filter for ECG signals", in Proc. IEEE Instrumentation and Measurement Technology Conference, Como, Italy, 2004, pp. 1706-1711.
[37] A.R. Verma, Y. Singh, "Adaptive Tunable Notch Filter for ECG Signal Enhancement", Procedia Computer Science 57, pp. 332-337, 2015.
[38] E. Ebrahimzadeh, M. Pooyan and et al., "ECG signals noise removal: Selection and optimization of the best adaptive filtering algorithm based on various algorithms comparison", Biomedical Engineering Applications Basis and Communications vol. 27 no 4, pp. 1-13, July, 2015.
[39] P. Chen, M. Chang and et al., "Study of Using Fourier Transform to Capture the ECG Signals between Awakeness and Dozing", in Proc. IS3C, Xian, China, 2016, pp. 1055-1058.
[40] C.E. Shannon (1948), 'A mathematical theory of communication'. The Bell System Technical Journal, 27, pp. 379-426.
[41] G. Egnal (1999), 'Image registration using mutual information', Technical Report, University of Pennsylvania.